\documentclass[aps,pre,twocolumn,showpacs,floatfix,superscriptaddress]{revtex4-1}
\usepackage{subfigure,amsmath, amsthm, amssymb}
\usepackage{graphicx}
\usepackage{graphics}
\usepackage[usenames]{color}

\usepackage[geometry]{ifsym}

\newcommand{\be}{\begin{equation}}
\newcommand{\ee}{\end{equation}}
\newcommand{\bea}{\begin{eqnarray}}
\newcommand{\eea}{\end{eqnarray}}

\begin{document}
\title{Shock propagation following an intense explosion: comparison between hydrodynamics and simulations}
\author{Jilmy P. Joy}
\email{jilmyp@imsc.res.in}
\affiliation{The Institute of Mathematical Sciences, CIT Campus, Taramani, 
Chennai 600113, India}
\affiliation{Homi Bhabha National Institute, Training School Complex, Anushakti Nagar, Mumbai 400094, India}
\author{Sudhir N. Pathak}
\email{snpathak@ntu.edu.sg}
\affiliation{Division of Physics and Applied Physics, School of Physical and Mathematical Sciences, Nanyang Technological University, 
Singapore}
\author{R. Rajesh}
\email{rrajesh@imsc.res.in}
\affiliation{The Institute of Mathematical Sciences, CIT Campus, Taramani, 
Chennai 600113, India}
\affiliation{Homi Bhabha National Institute, Training School Complex, Anushakti Nagar, Mumbai 400094, India}
  
\date{\today}
\pacs{05.20.Jj, 47.40.Nm, 05.20.Dd}

\begin{abstract}
The solution for the radial distribution of pressure, density, temperature and flow velocity fields in a blast wave  propagating through a medium at rest, following an intense explosion, starting from  hydrodynamic equations, is one of the classic problems in gas dynamics. However, there is very little direct verification of the theory and its assumptions from simulations of microscopic models. In this paper, we compare the results and assumptions of the hydrodynamic theory with results from large scale event driven molecular dynamics simulations of a hard sphere gas in three dimensions. We find that the  predictions for the radial distribution of the thermodynamic quantities  do not match well with the numerical data. We improve the theory by replacing the ideal gas law with a more realistic virial equation of state for the hard sphere gas.  While this improves the theoretical predictions, we show that they still fail to describe the data well. To understand the reasons for this discrepancy, the different assumptions of the hydrodynamic theory are tested  within the simulations. A key assumption of the theory is the existence of a local equation of state. We validate this assumption by showing that the local pressure, temperature and density obey the equation of state for a hard sphere gas. However, the probability  distribution of the velocity fluctuations has non-gaussian tails, especially away from the shock front, showing that the assumption of local equilibrium is violated. This, along with neglect of heat conduction, could be the possible reasons for the mismatch between theory and simulations.
\end{abstract}

\maketitle

\section{\label{introduction} Introduction}

The hydrodynamic description of a blast wave following the sudden release of a large amount of energy in a localized region, like in a nuclear explosion,  is one of the classic problems in gas dynamics. Due to the input temperature being much higher than the ambient temperature, the gas expands outwards. This results in radially dependent pressure, density and temperature fields that are discontinuous at a moving shock front that separates the ambient gas from the perturbed gas. In the case of strong shocks, the perturbed matter moves faster than the rate at which energy is transferred through heat or sound modes, and the expansion of the gas is self-similar in time. From the global conservation law for energy, it is straightforward to obtain, using dimensional analysis,
that the radius of the shock front $R(t)$ scales with time $t$ as $R(t) \sim t^{2/d+2}$ where $d$ is the spatial dimension~\cite{gtaylor1950,gtaylor1950_2,jvneumann1963cw,lsedov_book,sedov1946}. In addition to the scaling law for $R(t)$, from the local conservation laws for density, momentum, and energy, it is possible to obtain self-similar scaling 
solution for the radial distribution of pressure, density, velocity, and temperature. The exact solution was found by Taylor, von Neumann and Sedov~\cite{gtaylor1950,gtaylor1950_2,jvneumann1963cw,lsedov_book,sedov1946},  and will be referred to in the remainder of the paper as TvNS solution or TvNS theory. 

The scaling law for $R(t)$ was first verified, to a high degree of accuracy, in the Trinity nuclear explosion of 1945~\cite{gtaylor1950,gtaylor1950_2},  later on in the intermediate time evolution of supernova remnants~\cite{lsedov_book,lwoltjer1972,sfgull1973,dfcioffi1988}, and in laser-driven blast waves in gas jets~\cite{mjedwards2001prl}, plasma~\cite{adedensphysplasma2004}, or in cluster media~\cite{asmoorephysplasma2005}. Experimental studies and applications for evolution of astrophysical systems like supernova explosion are summarized in Ref.~\cite{jpostriker1988rmp,ybzeldovich_book}.

The TvNS theory also applicable to related problems. In some physical systems,  there is a continuous  input of energy in a localized region. 
The TvNS theory has been extended to  describe such cases both analytically~\cite{dokuchaev2002aanda} as well as numerically~\cite{saegfalle1975astronandastrophys}, and is relevant for   early stage of supernova explosions, powerful wind from stars and hidden neutrino sources~\cite{vsberezinsky2001astropartphys}, and for interstellar bubbles~\cite{jcastor1975apj,rweaver1977apj} . Other generalizations include the effect of including heat conduction~\cite{afghoniem1982jfm,amraouf1991fluiddynres,hsteiner1994physfluids}, viscous effect~\cite{neumann1950japplphys,rlatter1955japplphys,hlbrode1955japplphys,mnplooster1970physfluids} and implosions~\cite{guderley1942,rblazarus1981siam,Whithambook,thirschler2003fluiddynres,ponchaut2006fluidmech,rkanand2013wavemotion}.
The TvNS theory is also of  interest in understanding the response of an inelastic system, for example granular systems, to localized perturbations, either as an impact or continuous in time. Examples of such systems include crater formation in a granular bed following an impact of an object or a continuous jet~\cite{amwalsh2003prl,ptmetzger2009aip,ygrasselli2001granmatt}, shock propagation in a granular medium following a sudden impact~\cite{jfboudet2009prl,zjabeen2010epl,snpathak2012pre}, viscous fingering by the continuous injection of energy~\cite{xcheng2008natphy,bsandnes2007prl,sfpinto2007prl,ojohnsen2006pre,hhuang2012prl}, shock propagation in continuously driven granular media~\cite{jpjoy2017pre}, etc. The conservation laws for such systems are fewer in number, as energy is not conserved, and the TvNS theory has recently 
been generalized to include dissipative interactions~\cite{mbarbier2015prl,mbarbier2016physfluids}.

While the hydrodynamic equations and their modifications have been studied in detail, there has been little or no verification of the theory or its assumptions using microscopic models. It is only been more recently that the TvNS theory has been tested in simulations of microscopic models in which kinetic energy is given to a few particles at the center of a collection of stationary elastic particles. The radius of the disturbance has been shown to match with the TvNS prediction in both two~\cite{tantal2008pre,zjabeen2010epl} and three dimensions~\cite{zjabeen2010epl}. The predictions for the radial distribution of density, flow velocity and temperature fields in two dimensions have recently been compared with results from molecular dynamics simulations~\cite{mbarbier2016physfluids}. Here, in the   TvNS solution, the ideal gas law was replaced with  a more realistic constitutive relation expressing pressure in terms of density and temperature.  It was found that the simulations reproduce well the TvNS solution for low to medium densities, except for a small difference in the discontinuities at the shock front, and a slight discrepancy near the shock center. When the number density of the ambient gas is high, the TvNS solution was seen to not describe well the data near the shock front~\cite{mbarbier2016physfluids}. However, the key assumptions of local equilibrium and the existence of an equation of state that are assumed in the TvNS solution were not tested in the simulations. Also, the simulation data was presented only for a single time, and thus it is not very clear whether the scaling limit has been reached. While these simulations were in two dimensions, there are no similar studies in three dimensions. Thus, it is not very clear where the TvNS solution is reproducible in microscopic models and which assumptions of the TvNS theory are actually valid.

In this paper, we perform extensive event driven molecular dynamics simulations of hard spheres in three dimensions to test the predictions as well as the validity of the assumptions of the TvNS theory. We show unambiguously that the TvNS theory fails to describe the numerical data for most distances, ranging from the shock center to the shock front. We modify the TvNS theory to replace the constitutive relation from the ideal gas law to the virial equation of state (EOS) (up to 10 terms) for a hard sphere gas. While inclusion of the more realistic constitutive relation into the TvNS theory modifies  the predictions for the scaling functions, especially near the shock front, they still fail to describe the data well in terms of the exponents characterizing the power law behavior near the shock center. To understand this discrepancy, we  test numerically the various assumptions of the TvNS theory. A key assumption of the TvNS theory is the existence of an equation of state linking pressure, temperature and density. By measuring these quantities  independently, we show that numerically, the virial EOS is satisfied, justifying this key assumption. We also find that energy is equipartitioned equally among the different degrees of freedom, as would be expected in a system in local equilibrium.  However, we find that the  distribution of the velocity fluctuations, in regions between the shock center and shock front, has non-gaussian tails. In particular, it is asymmetric with non-zero skewness and an exponential tail, showing that local thermal equilibrium is not reached. The lack of local equilibrium could be a possible reason for the TvNS theory to fail in three dimensions. Also, we check that within the densities that we have studied, the motion is subsonic within the wave and supersonic with respect to the ambient gas, as implicitly assumed in the TvNS solution. Effects of including heat conduction are included as a discussion.

The remainder of the paper is organized as follows. In Sec.~\ref{analytical-numerical}, we briefly review the TvNS solution of the shock problem. In Sec.~\ref{comparison}, we give details of the simulations and  we present a detailed comparison between the numerically obtained data and the TvNS solution for two different number densities of the ambient gas.  In Sec.~\ref{virial}, we describe how the TvNS theory is modified when a more realistic virial EOS is used for the hard sphere gas. The effect of including higher order terms is discussed and compared with results from simulation. In  Sec.~\ref{local-equilibrium}, we show that, within the simulations, the virial EOS is obeyed locally. Also, equipartitioning of energy is shown to hold. It is also shown that the distribution of the velocity fluctuations is non-gaussian. In Sec.~\ref{soundvelocity}, we show that the flow velocity is subsonic within the blast and supersonic when compared to the ambient gas into which the shock is expanding. We conclude with a summary and discussion of our results in Sec.~\ref{conclusion}.

\section{\label{analytical-numerical} Review of T\lowercase{v}NS solution}

In this section, we briefly review the TvNS solution for the propagation of shock following an intense, isotropic, localized explosion. The equations of motion for a compressible fluid is obtained from the conservation laws for density, momentum, and energy. Due to isotropy, these quantities depend only on the radial distance $r$. Assumptions of local equilibrium, and  absence of heat conduction simplify the equations in the hydrodynamic limit. In particular, the assumption of local equilibrium implies that the flow is isentropic, and  that the local temperature, pressure and density are related through the EOS of the gas. In the TvNS solution, the EOS  is assumed to be that of the ideal gas. The resulting equations for the conservation laws are~\cite{Whithambook,Landaubook,barenblatbook}
\begin{align}
&\partial_t\rho+\partial_r(\rho v)+ 2 r^{-1}\rho v=0, \label{eq:conservation-density} \\ 
&\partial_t v+v\partial_r v+ \rho^{-1}\partial_r p=0,  \label{eq:conservation-momentum}\\ 
&\partial_t(p \rho^{-\gamma})+v\partial_r(p \rho^{-\gamma})=0,  \label{eq:conservation-entropy}
\end{align}
where $\rho(r,t)$ is the density, $v(r,t)$ is the radial velocity, $p(r,t)$ is the pressure, $\gamma$ is the adiabatic index, and $r$ is the radial distance from the location of the initial disturbance. Equations~(\ref{eq:conservation-density})--(\ref{eq:conservation-entropy}) describe the conservation of mass,  momentum, and entropy [for an ideal gas, entropy is proportional to $p\rho^{-\gamma}$] respectively.  

Non-dimensionalising the different thermodynamic quantities converts the Eqs.~(\ref{eq:conservation-density})--(\ref{eq:conservation-entropy}) into ordinary differential equations. From dimensional analysis,
\bea
p&=&\frac{\rho_0 r^2}{t^2}P(\xi), \nonumber \\
\rho&=&\rho_0 R(\xi), \label{eqn:dimensional_analysis} \\
v&=&\frac{r}{t}V(\xi), \nonumber \\
\varepsilon&=& \frac{k_B T}{m_0} = \frac{r^2}{t^2}E(\xi), \nonumber
\eea
where 
\be
 \xi=r\left(\frac{E_0 t^2}{\rho_0}\right)^{-1/5},
 \label{eqn:independent_variable}
\ee
is non-dimensionalised length, $E_0$ is the initial energy that is deposited into the system, $\rho_0$ is the ambient mass density, $T$ is the local temperature, $k_B$ is Boltzmann constant, $m_0$ is the mass of a particle, and $P$, $R$, $V$, and $E$, are scaling functions. $\varepsilon$ is the thermal energy per unit mass. The four scaling functions are not independent, but related through the ideal gas law as 
\be
P = R E.
\label{eq:relation}
\ee
In terms of the scaling functions, Eqs.~(\ref{eq:conservation-density})--(\ref{eq:conservation-entropy}) simplify to 
\begin{align}
& \left[V-\frac{2}{5}\right]\frac{RdV}{d\ln\xi}+\frac{dP}{d\ln\xi}-RV+RV^2+2P=0,\nonumber\\
& \frac{dV}{d\ln\xi}+\left[V-\frac{2}{5}\right]\frac{d\ln R}{d\ln\xi}+3V=0,\label{eqn:ode's}\\
& \frac{d}{d\ln\xi}\left(\ln\frac{P}{R^\gamma}\right)-\frac{2(1-V)}{V-2/5}=0.\nonumber
\end{align}

The thermodynamic quantities are discontinuous across the shock front, the discontinuities being given by the Rankine-Hugoniot conditions~\cite{Whithambook,Landaubook}. In terms of the scaling functions, these discontinuities are given by: 
\be
 P(\xi_f) = \frac{8}{25(\gamma+1)},
 V(\xi_f) = \frac{4}{5(\gamma+1)}, R(\xi_f) = \frac{\gamma+1}{\gamma-1},  
 \label{eqn:odes's_boundary_conditions}
\ee
where $\xi_f$ is the position of the shock front. $\xi_f$ is determined by the condition that total energy is conserved:
\be
4 \pi \int_0^{\xi_f}R(\xi)\left[\frac{V^2(\xi)}{2}+\frac{P(\xi)}{(\gamma-1)R(\xi)}\right]\xi^4 d\xi = 1.
\label{eqn:ode's_constraint}
\ee

The exact solution of Eq.~(\ref{eqn:ode's}) in three dimensions with the boundary condition as in Eq.~(\ref{eqn:odes's_boundary_conditions}) is~\cite{gtaylor1950,jvneumann1963cw,lsedov_book,sedov1946,barenblatbook}: 
\begin{align}
& \left[\frac{\xi_f}{\xi}\right]^5=C_1 V^2 \left[1-\frac{3\gamma-1}{2}V\right]^{\nu_1}\left[\frac{5}{2}\gamma V-1\right]^{\nu_2} \label{eq:exactV},\\
& R=C_2 \left[\frac{5}{2}\gamma V-1\right]^{\nu_3}\left[1-\frac{3\gamma-1}{2}\right]^{\nu_4}\left[1-\frac{5}{2}V\right]^{\nu_5} \label{eq:exactR},\\
&P=C_3 V^2\left[1-\frac{5 V}{2} \right]^{\nu_6}\left[\frac{5 \gamma V}{2} -1\right]^{\nu_7}\left[1-\frac{(3\gamma-1)V}{2}\right]^{\nu_8},
\label{eq:exactP}
\end{align}
where
\bea
 C_1&=&\left[\frac{5}{4}(\gamma+1)\right]^2\left[\frac{5(\gamma+1)}{7-\gamma}\right]^{\nu_1}\left(\frac{\gamma+1}{\gamma-1}\right)^{\nu_2},\nonumber\\
 C_2&=&\left(\frac{\gamma+1}{\gamma-1}\right)^{\nu_3+\nu_5+1}\left[\frac{5(\gamma+1)}{7-\gamma}\right]^{\nu_4},\\
 C_3&=&\frac{1}{2}\left(\frac{\gamma+1}{\gamma-1}\right)^{-\nu_5-\nu_3}(1+\gamma)\left[\frac{5(1+\gamma)}{7-\gamma}\right]^{\nu_4},\nonumber\\
 \label{eqn:power1}
 \eea
and
 \begin{align}
 \nu_1&=\frac{12-7\gamma+13\gamma^2}{-1+\gamma+6\gamma^2},
&  \nu_2&=-\frac{5(\gamma-1)}{2\gamma+1},\nonumber\\
  \nu_3&=\frac{3}{1+2\gamma},
& \nu_4&=\frac{-12+7\gamma-13\gamma^2}{(-2+\gamma)(-1+\gamma+6\gamma^2)},\nonumber\\
 \nu_5&=\frac{2}{-2+\gamma},
&  \nu_6&=\frac{\gamma}{-2+\gamma},\nonumber\\
  \nu_7&=\frac{2-2\gamma}{1+2\gamma},
&  \nu_8&=\nu_4. 
 \end{align}

Near the shock front, the four scaling functions $V$, $R$, $P$, and $E$ are discontinuous, as given by Eq.~(\ref{eqn:odes's_boundary_conditions}). Near the shock center ($\xi \to 0$), the scaling functions have power law singularities. These may be determined from Eqs.~(\ref{eq:exactV}) to (\ref{eq:exactP}). For monoatomic gas, as in our simulations, $\gamma=5/3$. From Eq.~(\ref{eq:exactV}), when $\xi \to 0$, it is easy to see that $V \to 2/(5 \gamma)$. $V$ approaches the limit as $V-2/(5 \gamma) \sim \xi^{-5/\nu_2}$ [$\xi^{13/2}$  for $\gamma=5/3$]. From Eqs.~(\ref{eq:exactR}) and (\ref{eq:exactP}), we obtain $R(\xi)\sim\xi^{-5 \nu_3/\nu_2}$ [$\xi^{9/2}$  for $\gamma=5/3$], and $P(\xi)\sim\xi^{-5 \nu_7/\nu_2}$ [$\xi^{-2}$  for $\gamma=5/3$]. Finally, from the ideal gas law $P=RE$, we obtain that the scaled thermal energy $E \sim \xi^{-5(\nu_7- \nu_3)/\nu_2}$ [$\xi^{-13/2}$  for $\gamma=5/3$]. Summarizing, for $\gamma=5/3$, when $\xi \to 0$,
\be
V-\frac{6}{25} \sim \xi^{13/2}, R(\xi)\sim\xi^{9/2}, P(\xi)\sim \xi^{-2}, E \sim \xi^{-13/2}.
\label{eq:results-summary}
\ee

\section{\label{comparison} Comparison with event- driven simulations}

In this section, we compare the predictions of the TvNS solution for density, velocity, pressure, and temperature profiles with results from large scale event-driven simulations~\cite{dcrapaportbook} of a collection of elastic hard spheres.  Consider a system of identical hard spheres in $3$ dimensions whose mass and diameter are set to one. Initially, all particles are at rest and uniformly distributed in space. The system is  perturbed by an initial localized input of energy at the origin. We model an isotropic impulse by giving a speed $v_0=1$ to $6$ particles near the center along $\pm x$, $\pm y$ and $\pm z$ directions. The particles move ballistically until they undergo elastic momentum conserving collisions with other particles. If $\vec{u}{_1}$ and $\vec{u}{_2}$ are the velocities of two particles $1$ and $2$ before collision, then the velocities after collision, $\vec{v}{_1}$ and $\vec{v}{_2}$, are given by
\bea
\vec{v}{_1} &=& \vec{u}{_1} -  [\hat{n} \cdot (\vec{u}{_1}-\vec{u}{_2})] \hat{n}, \nonumber \\
\vec{v}{_2} &=& \vec{u}{_2} -  [\hat{n} \cdot (\vec{u}{_2}-\vec{u}{_1})] \hat{n},
\eea
where $\hat{n}$ is the unit vector along the line joining the centers of particles $1$ and $2$. 

We simulate systems with number densities $\rho_0= 0.1$ and $0.4$, much smaller than the random closed packing density.  The system is  large enough so that the moving particles do not reach the boundary up to the times we have simulated, so that there are no boundary effects. The data are averaged over $150$ different histories. We also check that the data for an intermediate density [$\rho_0=0.25$] show a trend that is intermediate between that for $\rho_0= 0.1$ and $0.4$.

The initial perturbation creates a disturbance that propagates outwards in a radially symmetric fashion. A shock front separates the moving particles from the stationary ones. The radius of this shock front has been earlier shown to increase as $t^{2/5}$ in event driven simulations, consistent with the TvNS solution~\cite{zjabeen2010epl}.

Our aim is to numerically determine the scaling functions $R(\xi)$, $V(\xi)$, $E(\xi)$, and $P(\xi)$, corresponding to density, velocity,  temperature, and pressure respectively. The thermal energy is measured from the velocity fluctuations obtained by subtracting out the mean radial velocity $v_r (r, t) \hat{r}$ from the instantaneous velocity. The pressure is measured from the local collision rate. For the hard sphere gas in three dimensions,  pressure is given by the expression~\cite{masaharumdsimulation2016} 
\be
 p= \rho T - \frac{\rho}{3 N \delta t} \sum_{collisions}b_{ij},
 \label{eqn:pressure}
\ee
where $b_{ij}=\vec{r}_{ij} \cdot \vec{v}_{ij}$, where $\vec{r}_{ij}$ and $\vec{v}_{ij}$ respectively are the relative positions and velocities of the particles $i$ and $j$ undergoing collisions, $\delta t$ is the time duration of measurement, and $N$ is the mean number of particles in the radial bin whose pressure is being computed.

\begin{figure}
 \includegraphics[width=\columnwidth]{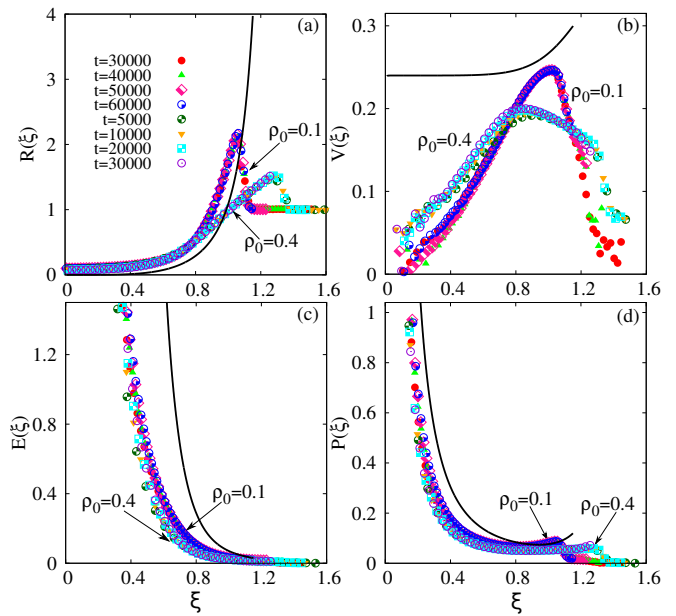}
 \caption{(Color online)  The variation of the scaling functions (a) $R(\xi)$, (b) $V(\xi)$, (c) $E(\xi)$  and (d) $P(\xi)$ corresponding to non-dimensionalised density, velocity, temperature and pressure  [see Eq.~(\ref{eqn:dimensional_analysis})] with scaled distance  $\xi$. The data are shown for $2$ different initial densities $\rho_0 =0.1$ and $0.4$. For $\rho_0=0.1$, the different times are $t=30000$, $40000$, $50000$, $60000$, and for $\rho=0.4$, $t=5000$, $10000$, $20000$, $30000$,  as indicated in (a).  The black solid lines correspond to the TvNS solution [see Eqs.~(\ref{eq:exactV})-(\ref{eq:exactP}) and Eq.~(\ref{eq:relation})]. The data for $R$, $P$, and $E$ are also shown on a logarithmic scale  in Fig.~\ref{fig:data_analysis_comparison_powerlaw}.}
 \label{fig:data_analysis_comparison}
\end{figure}

The numerically obtained scaling functions $R(\xi)$, $V(\xi)$, $E(\xi)$, and $P(\xi)$ are shown in Fig.~\ref{fig:data_analysis_comparison} for initial number  densities $0.1$ and $0.4$. For each of the densities, four different times are shown. The data for the different times collapse onto one curve when plotted against $\xi$. The TvNS solution is shown in solid black line. Since the simulations correspond to a monoatomic gas, we set $\gamma=5/3$ in the TvNS solution.

\begin{figure}
 \includegraphics[width=0.9\columnwidth]{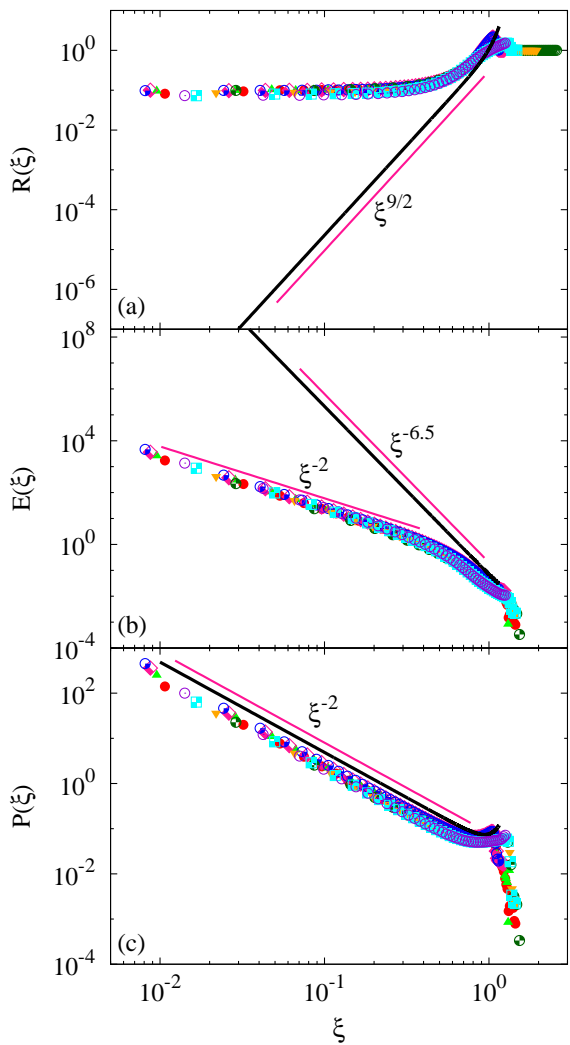}
 \caption{(Color online)  The data in Fig.~\ref{fig:data_analysis_comparison}(a), (c) and (d) are shown in logarithmic scale to emphasize the power-law divergence for small $\xi$. The three plots show the variation of  the scaling functions (a) $R(\xi)$, (b) $E(\xi)$  and (c) $P(\xi)$ with scaled distance  $\xi$. The data are for $2$ different initial densities $\rho_0 =0.1$ and $0.4$. Each density has data for four different times and the symbols are same as described in Fig~\ref{fig:data_analysis_comparison} (a). The black solid lines correspond to the TvNS solution.}
 \label{fig:data_analysis_comparison_powerlaw}
\end{figure}

The scaling function $R(\xi)$ depends on the initial number density $\rho_0$, especially close to the shock front [see Fig~\ref{fig:data_analysis_comparison}(a)]. As $\rho_0$ decreases, the discontinuity in density at the shock front increases, and it is possible that it may approach the theoretical result in the limit of infinite dilution. However, when compared with the entire range of $\xi$, the numerically obtained curves are very different from the TvNS solution. In particular, as shown in Fig.~\ref{fig:data_analysis_comparison_powerlaw}(a), the TvNS scaling function increases as a power law $\xi^{9/2}$ for small $\xi$ [as in Eq.~(\ref{eq:results-summary})], while the numerically obtained scaling function tends to a non-zero constant. 

The scaling function $V$, shown in Fig~\ref{fig:data_analysis_comparison}(b), increases linearly from zero, reaches a maximum and then decreases to its value at the shock front. The  data for $V$ from the simulations are again not consistent with the TvNS solution in which $V$ is non-zero at $\xi=0$, and then monotonically increases to its value at the shock front. Decreasing the ambient number density $\rho_0$ shifts the numerical data further away from the TvNS solution for small $\xi$.

The scaling function $E(\xi)$, which measures the local velocity fluctuations, is shown in Fig~\ref{fig:data_analysis_comparison}(c). There is only a weak  dependence on the number density $\rho_0$. From Fig~\ref{fig:data_analysis_comparison_powerlaw}(b), it can be seen that $E$ diverges  as a power law as $\xi \to 0$, with an exponent that we numerically estimate to be close to $-2$. The TvNS solution predicts that $E$ diverges with decreasing $\xi$ as $E\sim \xi^{-13/2}$ [see Eq.~(\ref{eq:results-summary})]. Again, the data from simulations are quantitatively different from the TvNS solution.

The dependence of the scaled pressure on $\xi$ is shown in
Fig~\ref{fig:data_analysis_comparison}(d). Unlike the other scaling
functions, the data from simulations are much closer to the TvNS
solution. In particular, both data diverge as $\xi^{-2}$ for small $\xi$
[see Fig~\ref{fig:data_analysis_comparison_powerlaw}(c)]. Also, as the
number density $\rho_0$ is decreased, the data tends towards the
analytical solution, though it is not possible to extrapolate the data to $\rho_0=0$ with the current data.

In summary, the  TvNS solution fails  to describe the numerical data. Also, the data from simulations show that the results depend on $\rho_0$. On the other hand the TvNS solution is independent of $\rho_0$. There could be multiple plausible reasons for the observed differences. One is that the simulations are performed for hard spheres, while the TvNS solution is for the ideal gas. In the hard sphere simulations, there is an upper bound for the local number density, while there is no such bound in the ideal gas. This particularly becomes significant near the shock front where the density becomes high irrespective of the initial density. Secondly, the medium through which shock is propagating is inherently a system out of equilibrium, and the assumption of local equilibrium in the TvNS solution may be incorrect. 

In the following, we test these assumptions. First, we incorporate effects of explosion in the hard sphere model into the TvNS solution by replacing the EOS for an ideal gas with a virial expansion for the hard sphere gas. Second, we test the assumption of local equilibrium.

\section{\label{virial} Shock propagation in a hard sphere gas}

The EOS and the free energy of the hard sphere gas may be expressed as a virial expansion:
\begin{align}
&\frac{p}{k_B T \rho}= 1+ \sum_{n=2}^{\infty} B_n \rho^{n-1}, \label{eqn:virial} \\
& \frac{F(N,V,T)}{Nk_B T}= \ln(\Lambda^d \rho)-1+\sum_{n=2}^{\infty}B_n \frac{\rho ^{n-1}}{n-1},
\end{align}
where $p$ is pressure, $T$ is temperature, $k_B$ is Boltzmann constant, $\rho$ is number density, $B_n$ is the $n$th virial coefficient, $F$ is the free energy, and $\Lambda = h/\sqrt{2 \pi m k_B T}$ is the thermal wavelength. For a hard sphere gas, the virial coefficients are independent of the temperature. 
Now, the set of hydrodynamic equations describing the propagation of the shock in the ideal gas [see Eqs.~(\ref{eq:conservation-density})-(\ref{eq:conservation-entropy})], have to be modified. It is more convenient to work in terms of the local temperature rather that the local pressure. Equations~(\ref{eq:conservation-density}) and (\ref{eq:conservation-momentum}) for density and momentum remain unchanged except for replacing pressure $p$ with the virial expansion in Eq.~(\ref{eqn:virial}). Equation~(\ref{eq:conservation-entropy}) for entropy is modified to
\be
(\partial_t + v \partial_r)  \frac{3 \ln T}{2} -\left[ 1\! +\! \sum_{n=2}^{\infty} B_n \rho^{n-1} \right]\!(\partial_t + v \partial_r)  \ln \rho = 0.
  \label{eqn:virial_conservation_laws}
\ee

Transforming to dimensionless variables using Eqs.~(\ref{eqn:dimensional_analysis}), the hydrodynamic equations reduce to
\begin{widetext}
\bea
0&=&  \left(V-\frac{2}{5}\right) \xi R \frac{dV}{d\xi} + \xi  \frac{d}{d\xi} \left[ER \left(1+ \sum_{n=2}^{\infty} B_n \rho_0^{n-1} R^{n-1} \right)\right]
 - R V + R V^2+ 2 R E\left[1 + \sum_{n=2}^{\infty} B_n \rho_0^{n-1} R^{n-1}\right], \label{eq:22}\\
0&=&  \left(V-\frac{2}{5}\right) \xi \frac{dR}{d\xi} + \xi R \frac{dV}{d\xi} + 3 R V,\\
0&=&  -\Big(1 + \sum_{n=2}^{\infty} B_n \rho_0^{n-1} R^{n-1}\Big) \left(V-\frac{2}{5}\right) \frac{\xi}{R} \frac{dR}{d\xi} + \frac{3}{2} \left(V-\frac{2}{5}\right) \frac{\xi}{E} \frac{dE}{d\xi} +3 (V-1).
  \label{eqn:ode_virial}
\eea
\end{widetext}
The Rankine-Hugoniot boundary conditions at the shock front $\xi_f$ are now
\begin{align}
&\frac{1}{R(\xi_f )}\left[1+\frac{3}{1+ \sum_{n=2}^{\infty} B_n \rho_0^{n-1} R(\xi_f )^{n-1}}\right]=1,\nonumber \\
&V(\xi_f )= \frac{6}{5} \frac{1}{R(\xi_f )[1+ \sum_{n=2}^{\infty} B_n \rho_0^{n-1} R(\xi_f)^{n-1}]}, \nonumber\\
&E(\xi_f ) =\frac{1}{3} V(\xi_f )^2. \label{eq:rankine}
\end{align}
For a given $\xi_f$, Eqs.~(\ref{eq:22})-(\ref{eqn:ode_virial}) with the boundary conditions in Eqs.~(\ref{eq:rankine}) may be solved numerically. $\xi_f$, as before, is determined by the condition that total energy is conserved, which in terms of the scaling functions is
\be
4 \pi \int_0^{\xi_f}R(\xi)\left[\frac{V^2(\xi)}{2}+\frac{3}{2} E(\xi) \right]\xi^4 d\xi = 1,
\label{eqn:xif_constraint}
\ee
which is same as Eq.~(\ref{eqn:ode's_constraint}) with $\gamma=5/3$ and $P=RE$.

For the hard sphere gas in three dimensions, the virial coefficients $B_n$ are known analytically for up to $n=4$ and through Monte Carlo simulations up to $n=10$~\cite{McCoybook}. These are tabulated in  Table~\ref{table:Bn}.
\begin{table}
\caption{The values of the virial coefficients $B_n$ for the hard sphere gas in three dimensions. The data are taken from Ref.~\cite{McCoybook}.}
\begin{center}
\begin{tabular}{ c c }
 \hline\hline
 $n$ & $B_n$\\[0.5ex]
 \hline 
 2 &  $\frac{2\pi}{3}$\\
 $3$ & $ \frac{5}{8} B_2^2$\\
 $4$ & $\Big[\frac{2707}{4480}+\frac{219 \sqrt{2}}{2240 \pi}-\frac{4131}{4480}\frac{\arccos{[1/3]}}{\pi}\Big] B_2^3$\\ 
 $5$ & $0.110252 B_2^4$\\
 $6$ & $0.03888198 B_2^5$\\
 $7$ & $0.01302354 B_2^6$\\
 $8$ & $0.0041832 B_2^7$\\
 $9$ & $0.0013094 B_2^8$\\
 $10$ & $0.0004035 B_2^9$\\[1ex]
 \hline \hline
\end{tabular}
\end{center}
\label{table:Bn}
\end{table}

From the hydrodynamic equations as well as the boundary conditions, it is clear that the initial density $\rho_0$ now affects the results, since it explicitly appears in the equations. The limit $\rho_0 \to 0$ in these equations should reproduce the TvNS solution. We thus expect that there could be significant deviations for larger densities. We now compare the results for the solution for the hydrodynamic equation with virial EOS for hard sphere gas with that for the ideal gas as well as those obtained from simulations.

\begin{figure}
 \includegraphics[width=\columnwidth]{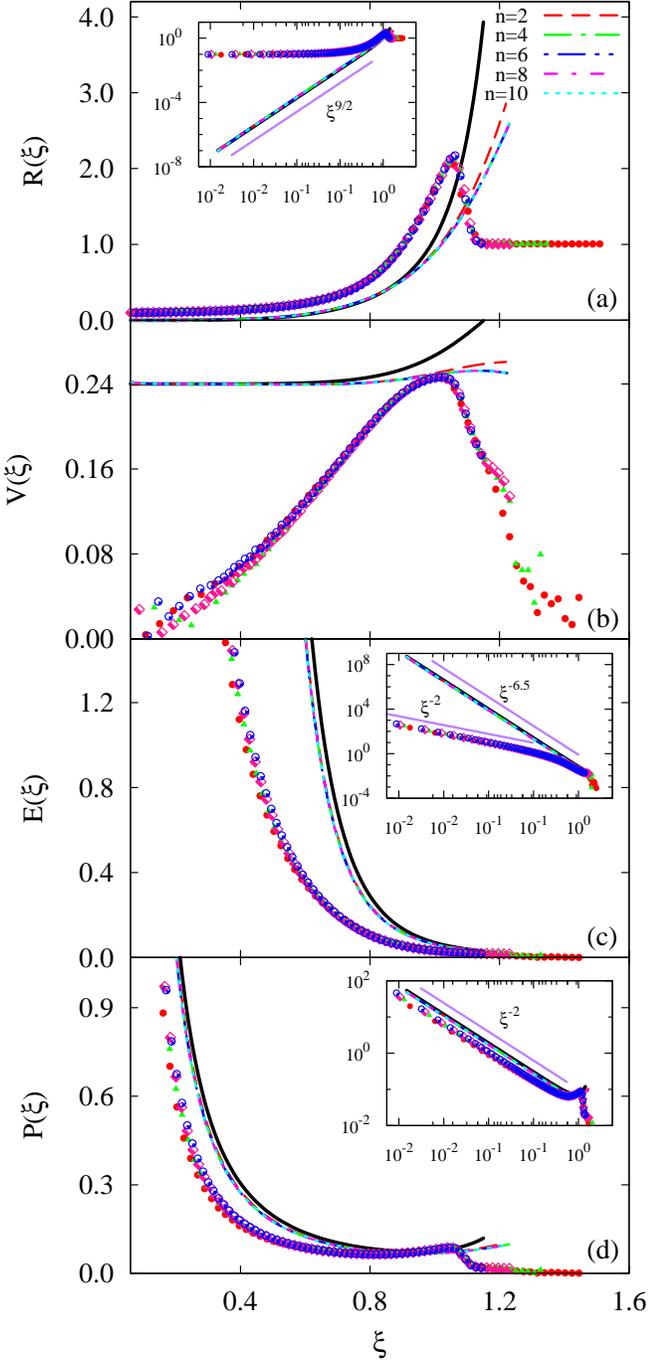}
 \caption{(Color online) The scaling functions (a) $R(\xi)$, (b) $V(\xi)$, (c) $E(\xi)$, and (d) $P(\xi)$ corresponding to density, velocity, temperature and pressure respectively versus $\xi$ for ambient density $\rho_0=0.1$ is compared with the theoretical solution for the hydrodynamic equations with virial EOS for the hard sphere gas. The simulation data (represented by points) correspond to four different times with keys as shown in Fig~\ref{fig:data_analysis_comparison}(a). The  lines represent the virial EOS solution with the virial expansion truncated at $n=2, 4, 6, 8, 10$.  Black solid curve represents the case of  ideal gas. The inset shows the plots on a log-log scale, accentuating the small $\xi$ behavior.} 
 \label{fig:virial-van_der_waals_rho_0.1}
\end{figure}

\begin{figure}
 \includegraphics[width=\columnwidth]{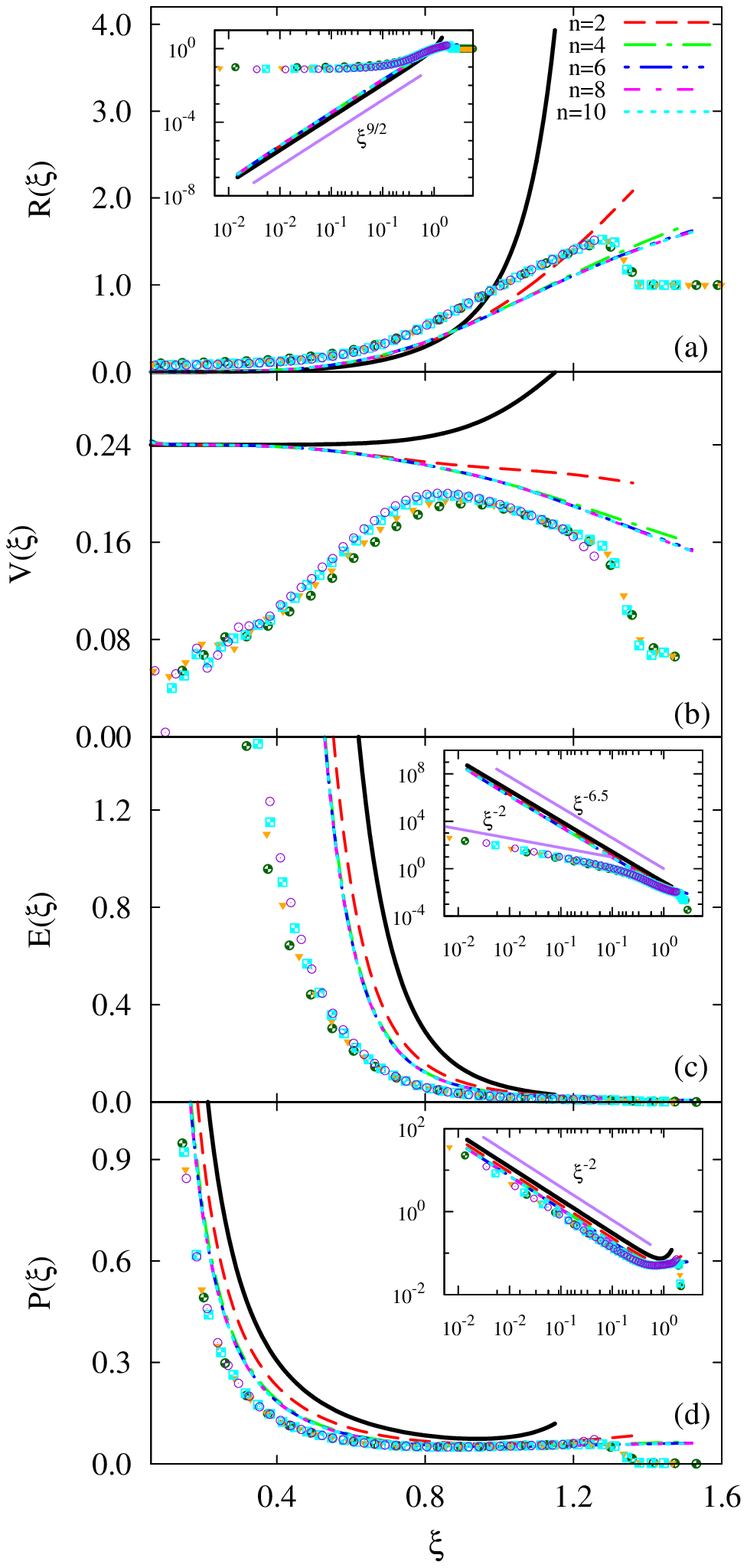}
 \caption{(Color online) The scaling functions (a) $R(\xi)$, (b) $V(\xi)$, (c) $E(\xi)$, and (d) $P(\xi)$ corresponding to density, velocity, temperature and pressure respectively versus $\xi$ for ambient density $\rho_0=0.4$ is compared with the theoretical solution for the hydrodynamic equations with virial EOS for the hard sphere gas. The simulation data (represented by points) correspond to four different times with keys as shown in Fig~\ref{fig:data_analysis_comparison}(a). The  lines represent the virial EOS solution with the virial expansion truncated at $n=2, 4, 6, 8, 10$.  Black solid curve represents the case of  ideal gas. The inset shows the plots on a log-log scale, accentuating the small $\xi$ behavior.} 
 \label{fig:virial-van_der_waals_rho_0.4}
\end{figure}

In Figs.~\ref{fig:virial-van_der_waals_rho_0.1} and \ref{fig:virial-van_der_waals_rho_0.4}, the TvNS solution with virial EOS is denoted by lines while the simulation data are shown by points for $\rho_0=0.1$ and $\rho_0=0.4$ respectively. We first focus on the effect of truncating the virial EOS by including only the first $n$ virial terms. We find that the scaling functions for density, velocity, temperature and pressure for $n=6,8,10$ are indistinguishable from each other for both densities [see Figs.~\ref{fig:virial-van_der_waals_rho_0.1} and \ref{fig:virial-van_der_waals_rho_0.4}], showing that errors introduced by truncating the EOS  at $n=10$ are negligible. Second, we observe that as $n$ or $\rho_0$ increases, the discontinuities at the shock front decreases, and the position of the shock front $\xi_f$ increases. Third, and more importantly, the inclusion of virial EOS does not alter the exponents of the power law divergence of the scaling functions at small $\xi$. Also, the exponents are independent of $\rho_0$. For high densities, there are qualitative changes induced by including virial EOS. For example, the scaling function for velocity  changes from monotonically increasing for ideal gas to monotonically decreasing for the virial EOS [see Fig.~\ref{fig:virial-van_der_waals_rho_0.4}(b)].

Near the shock front, in comparison to the ideal gas EOS, we find that the virial EOS does a better job of describing the scaling functions obtained from simulations. For the density scaling function, the  discontinuity at the shock front is better captured by the virial EOS for both densities [see Figs.~\ref{fig:virial-van_der_waals_rho_0.1}(a) and \ref{fig:virial-van_der_waals_rho_0.4}(a)].  For the velocity scaling function, the virial EOS matches with the simulation data close to the shock front for $\rho_0=0.1$ [see Fig.~\ref{fig:virial-van_der_waals_rho_0.1}(b)] , and captures the correct trend for $\rho_0=0.4$ [see Fig.~\ref{fig:virial-van_der_waals_rho_0.4}(b)]. For both temperature [see Figs.~\ref{fig:virial-van_der_waals_rho_0.1}(c) and \ref{fig:virial-van_der_waals_rho_0.4}(c)] and pressure [see Figs.~\ref{fig:virial-van_der_waals_rho_0.1}(d) and \ref{fig:virial-van_der_waals_rho_0.4}(d)], the scaling functions obtained from virial EOS are closer to the simulation data than the ideal gas EOS. 

Away from the shock front, the power law behavior of the scaling function remain unchanged with the inclusion of the virial EOS. This is because near the shock center, the gas is dilute, and the virial EOS tends towards the ideal gas law.  Thus, as for the TvNS theory discussed in Sec.~\ref{comparison}, the modified theory fails to describe the simulation data [see insets of Figs.~\ref{fig:virial-van_der_waals_rho_0.1} and \ref{fig:virial-van_der_waals_rho_0.4}].

Thus, we conclude that while replacing the ideal gas assumption in the TvNS solution with the virial EOS for the hard sphere gas introduces some dependence on the initial density $\rho_0$, it fails to capture the strong deviations from the TvNS solution that is observed in the event driven simulations.

\section{\label{local-equilibrium} Local Equilibrium}

We now numerically check the assumption of local equilibrium in the TvNS solution. First, we check whether the EOS of a hard sphere gas is obeyed. Second, we check whether the thermal energy is locally equipartitioned equally in all three directions. Finally, we calculate the skewness and kurtosis of the  distribution of velocity fluctuations to check the deviation from a gaussian.

\subsection{Equation of state}

A central assumption of the TvNS solution is that the local pressure, density and temperature are not independent, but related to each other through an EOS, which is assumed to be that of the ideal gas. To test the assumption of EOS, we measure the local thermodynamic quantities and check whether they obey the  hard sphere virial EOS. To do so, we measure the ratio
\be
\chi(n)=   \frac{P(\xi)}{E(\xi) R(\xi)\left[1 + \sum_{k=2}^{n} B_k \rho_0^{k-1} R(\xi)^{k-1}\right]},
 \label{eqn:scaling_fn_virial}
\ee
where $n$ is the number of terms retained in the virial expansion [$n=1$ corresponds to ideal gas]. For large $n$, if $\chi \approx 1$, then we conclude that the virial EOS is obeyed, and the assumption of EOS is justified.

\begin{figure}
 \includegraphics[width=\columnwidth]{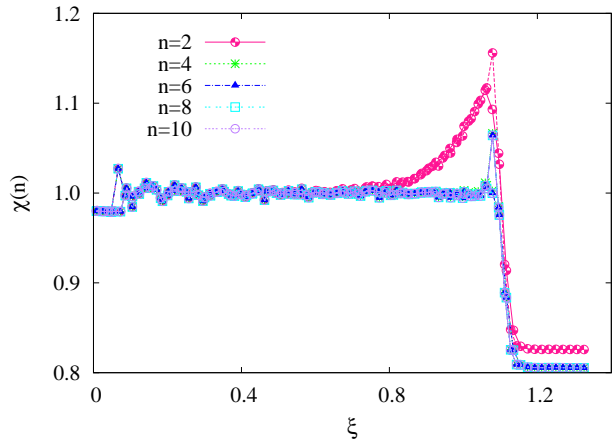}
 \caption{(Color online)  The variation of $\chi(n)$ [see Eq.~(\ref{eqn:scaling_fn_virial})] with $\xi$ for $n=2,4,6,8, 10$. The data are for  times $40000$ and $60000$ and for ambient number density $\rho_0=0.1$.  For large $n$, $\chi(n)$ fluctuates about $1$. }
 \label{fig:eqn_of_state}
\end{figure}

The dependence of $\chi(n)$ on $\xi$ for $n=2,4,6,8,10$ is shown in Fig.~\ref{fig:eqn_of_state}  for two different times. While for small $n$, there is deviation from one near the shock front, quite remarkably, as $n$ increases, $\chi(n)$ fluctuates about $1$ for all $\xi$. This clearly shows that the assumption of an EOS is quite justified.

\begin{figure}
 \includegraphics[width=\columnwidth]{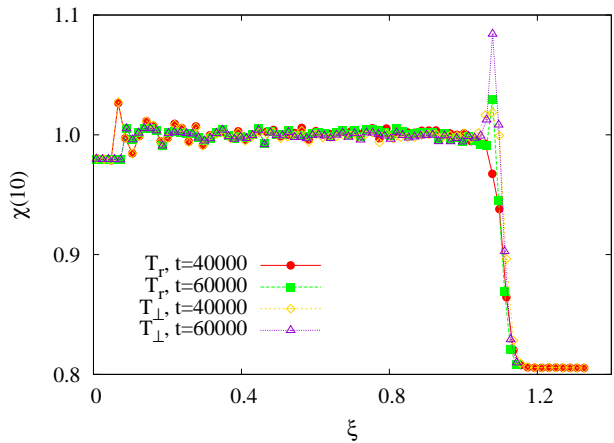}
 \caption{(Color online)  The variation of $\chi(10)$ with $\xi$, where the temperature in Eq.~(\ref{eqn:scaling_fn_virial}) is replaced by $T_{r}$ or $T_\perp$ defined through the radial and perpendicular components of the velocity fluctuations. The data are for times $40000$ and $60000$ and ambient density $\rho_0=0.1$.}
 \label{fig:eqn_of_state_Tr_Tperp}
\end{figure}

There is an ambiguity about whether we define temperature using the full velocity fluctuations or by only considering the radial or transverse components. To show that the evidence of EOS is not dependent on this choice, consider Fig~\ref{fig:eqn_of_state_Tr_Tperp}  where $\chi(10)$ is shown for both $T_{r}$ [based on radial component] and $T_\perp$ [based on transverse component]. Clearly, the results are independent of the choice, except very close to the shock front.

\subsection{Equipartition}

\begin{figure}
 \includegraphics[width=\columnwidth]{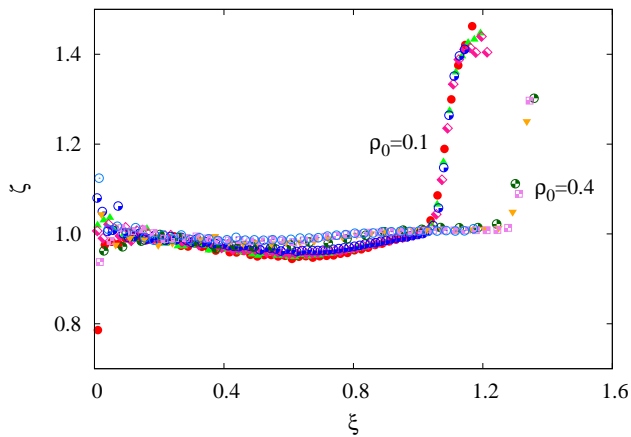}
 \caption{(Color online) The variation of $\zeta$, the ratio of energies in the radial and $\theta$-$\phi$ directions [see Eq.~(\ref{eq:zeta})] with the scaled distance $\xi$. The data is for four different times with keys as  in Fig~\ref{fig:data_analysis_comparison}(a), for two ambient densities $\rho_0=0.1$ and $0.4$. Away from the shock front, $\zeta \approx 1$.}
 \label{fig:equipartition}
\end{figure}

We check whether the thermal energy is equipartitioned equally in all three directions by measuring the ratio
\be
\zeta = \frac{ \langle \delta v_r^2 \rangle}{ \langle \delta v_\perp^2 \rangle/2},
\label{eq:zeta}
\ee
where $\delta v_r$ and $\delta v_\perp$ are the velocity fluctuations in the radial and transverse $\theta$-$\phi$ directions respectively. The factor of $2$ accounts for the two degrees of freedom in the $\theta$-$\phi$ directions. If the thermal energy is equipartitioned, then $\zeta=1$. Fig.~\ref{fig:equipartition} shows the variation of $\zeta$ with $\xi$ for different times. The data for different times collapse on to a single curve. Away from the shock front, $\zeta \approx 1$ showing equipartition. However, near the shock front, $ \zeta >1$, corresponding to excess thermal energy in the radial direction. 

\subsection{Skewness and Kurtosis}

\begin{figure}
 \includegraphics[width=\columnwidth]{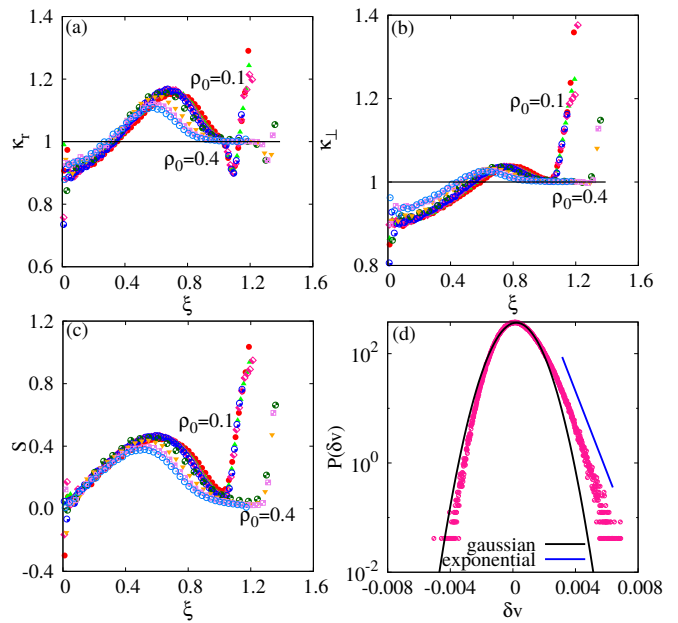}
 \caption{(Color online) The variation with scaled distance $\xi$ of (a) the kurtosis $\kappa_r$ for the radial velocity fluctuations [see Eq.~(\ref{eq:kurtosisr})]. (b) the  kurtosis $\kappa_\perp$ for the  velocity fluctuations in the $\theta$-$\phi$ direction [see Eq.~(\ref{eq:kurtosisperp})] and (c) skewness $S$ for the radial velocity fluctuations [see Eq.~(\ref{eq:skewness}].  The black solid line in (a) and (b) are  reference lines of $1$, corresponding to gaussianity. The data are for  $\rho_0 =0.1$ and $\rho_0=0.4$ and for  four different times  with keys as  in Fig~\ref{fig:data_analysis_comparison}(a). (d) The distribution of the radial velocity fluctuations $P(\delta v)$ measured at $r=61.5$, $t=30000$ and $\rho_0=0.4$, corresponding to $\xi=0.58$. The black solid curve represents the gaussian distribution fitted to the data near zero.  The blue solid line is an exponential and a guide to the eye.} 
 \label{fig:local_thermal_equilibrium}
\end{figure}

To quantify the deviation from gaussianity, we measure the kurtosis $\kappa$, and skewness $S$ of the probability distribution for the velocity fluctuations:
\bea
\kappa_r & = & \frac{\langle \delta v_{r}^4\rangle}{3 \langle \delta v_{r}^2\rangle^2}, \label{eq:kurtosisr}\\
\kappa_\perp & = &\frac{\langle \delta v_{\perp}^4\rangle}{2 \langle \delta v_{\perp}^2\rangle^2}, \label{eq:kurtosisperp}\\
S &=& \frac{\langle \delta v_r^3\rangle}{\langle \delta v_r^2\rangle^{3/2}}. \label{eq:skewness}
\eea
Deviation of kurtosis from $1$ shows non-gaussian behavior. Likewise, a non-zero skewness shows that the distribution is not symmetric. The variation of $\kappa_r$ and $\kappa_\perp$ with $\xi$ is shown in Fig.~\ref{fig:local_thermal_equilibrium} (a) and (b) respectively. While the data for different times collapse onto one curve, $\kappa_r$ and $\kappa_\perp$ deviate significantly from $1$ for almost all $\xi$, showing a lack of local equilibrium. For the higher density $\rho_0=0.4$, $\kappa_r$ and $\kappa_\perp$ are close to one near the shock front. Skewness $S$ also provides a strong evidence for deviation from gaussianity and 
a lack of local equilibrium. From the variation of $S$ with $\xi$, as shown in Fig.~\ref{fig:local_thermal_equilibrium} (c), it is clear that it is positive for all values of $\xi$. Thus, the distribution is clearly asymmetric. Note that in the $\theta$-$\phi$ directions, due to symmetry, the skewness is zero. 

The skewness of the distribution may be directly seen by examining the full probability distribution $P(\delta v_r, r,t)$ for the fluctuations in the radial velocity. The distribution for a fixed time $t$ and fixed radial distance $r$, corresponding to $\xi=0.58$ is shown in Fig~\ref{fig:local_thermal_equilibrium}(d). This corresponds to a region away from the shock front where the skewness in Fig.~\ref{fig:local_thermal_equilibrium}(c) is non-zero. It can be seen from the figure that the distribution deviates from the gaussian distribution and is skewed towards the larger positive fluctuations. We find that the data are consistent with an exponential decay for large positive fluctuations.

\section{\label{soundvelocity}Sonic line}

In this section, we check an implicit assumption of the TvNS theory. In the classic solution of a shock, the flow velocity is subsonic within the blast and supersonic when compared to the ambient gas into which the shock is expanding. The subsonic flow results in perturbations relaxing quickly. On the other hand, if there are regions where the flow is supersonic, then the asymptotic self similar solution may not be reached.
To check whether this assumption is valid both within the event driven simulations as well as the TvNS theory modified by the virial EOS, we follow 
closely the analysis of Ref.~\cite{mbarbier2016physfluids}.

Consider the local sound velocity,
\be
c(r,t) = \sqrt{\frac{\gamma p(r,t)}{\rho(r,t)}}.
\ee
Since a perturbation results in an acoustic wave (heat waves have been ignored), the energy carrying perturbation will have a maximal speed $v(r,t) + c(r,t)$, where $v(r,t)$ is the local flow velocity. On the other hand, the phase velocity  is the geometric speed at $r$ and is given by $\frac{r}{r_f} \frac{dr_f}{dt}$, where $r_f$ is the position of the shock front. Since $r_f \propto t^{2/5}$, we obtain  $\frac{dr_f}{dt} = \frac{2 r_f}{5 t}$, such that the phase velocity is $\frac{2r}{5 t}$. For the TvNS assumption to hold, the phase velocity should be less than the maximal speed $v(r,t) + c(r,t)$. This may be checked by plotting the scaled sound velocity $C(\xi)=t c/r$ as a function of scaled flow velocity $V(\xi)=tv/r$, and compare the data with sonic line
\be
 \frac{5}{2} V + \frac{5}{2} C =1.
 \label{eq:sonicline}
\ee
If the  $C-V$ plot stays above this line, then perturbations are faster than phase velocity, and is necessary for the validity of the TvNS solution.

\begin{figure}
 \includegraphics[width=\columnwidth]{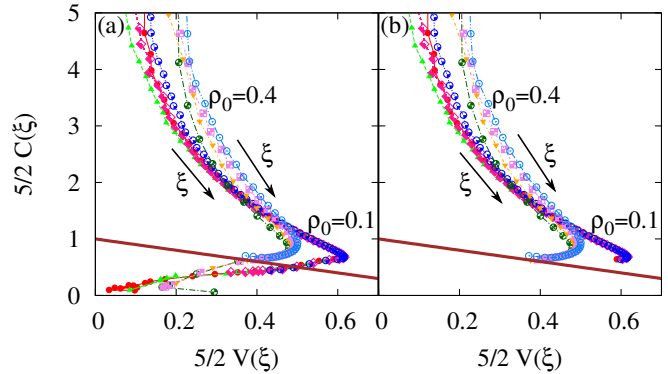}
 \caption{(Color online) The parametric plot of sound velocity(C)-flow velocity(V). The data are from simulations. The solid reference line represents the sonic line [see Eq.~(\ref{eq:sonicline}]. The arrows indicate the direction of increasing $\xi$. (a) The full data. (b) Reduced data, where the data beyond the shock front have been removed. The data are  for four different times with keys as in Fig~\ref{fig:data_analysis_comparison}(a).}
 \label{fig:sound_velocity}
\end{figure}

The parametric plot of the scaled sound velocity with the scaled flow velocity, as obtained from simulations, is shown in Fig~\ref{fig:sound_velocity}(a). While most of the data lies above the sonic line (shown as solid line), there is a region near the shock front where the data from simulation lies below the sonic line for both the two different ambient densities.  However, the region that we find below the sonic line may be an artifact of the simulation. The data is obtained by averaging over different histories, each one having a slightly different shock front, resulting in a diffused shock front. To remove this ambiguity, we identify  the value of $\xi$ at which density (or pressure) reaches a maximum as the shock front. We remove the data beyond the shock wave, and plot this reduced data in  Fig~\ref{fig:sound_velocity} (b). Here, we find that the entire plot lies above the sonic line.  So, we conclude that our simulations are in the parameter range corresponding to  strong shocks.

\begin{figure}
 \includegraphics[width=0.8\columnwidth]{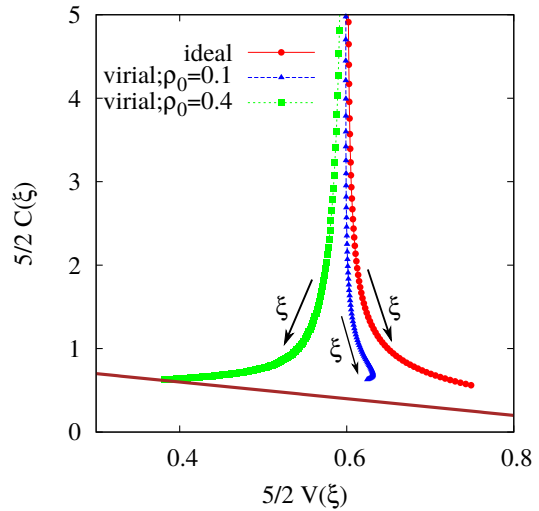}
 \caption{(Color online) The parametric plot of sound velocity(C)-flow velocity(V). The data are from TvNS solution with virial EOS truncated at the tenth virial coefficient. The solid reference line represents the sonic line [see Eq.~(\ref{eq:sonicline}]. The arrows indicate the direction of increasing $\xi$. }
 \label{fig:sound_velocity_virial}
\end{figure}

The same features may be observed for the theoretical TvNS solution with virial EOS. The C-V plot for the same is shown in Fig~\ref{fig:sound_velocity_virial}, where the virial EOS is truncated at the tenth virial coefficient. It may be seen that for the densities that we have considered, the C-V parametric plot lies above the sonic line. Any higher ambient density would have resulted in the curve crossing the sonic line.

\section{\label{conclusion} Conclusion and discussion}

To summarize, in this paper we revisited the classic solution describing the propagation of a blast wave through a medium at rest, following an intense explosion. We compared the TvNS solution for the radial distribution of pressure, temperature, density and flow velocity fields with results from large scale event driven molecular dynamics simulations of hard spheres in three dimensions. We find that the TvNS solution fails to describe the numerical data well. In particular, the power law behavior away from the shock front for temperature and density have different exponents in the theory and simulations. In addition, the predictions for the flow velocity do not match the simulation results for all $\xi$.

The TvNS theory was modified by using a virial equation of state for the hard sphere gas instead of the ideal gas constitutive relation. We find that the hydrodynamic solution does not noticeably change beyond the inclusion of six virial coefficients. We restricted our analysis to the known ten virial coefficients. While inclusion of the more realistic virial equation of state modifies the theoretical predictions, especially near the shock front, it does not modify any of the exponents, and thus fails to describe the simulation data. 

We also checked the different assumptions of the TvNS theory. The assumption of local equilibrium is a key assumption of the TvNS solution. In particular, the main consequence of this assumption that goes into the theory is the existence of an equation of state for the gas. Though the numerics and theory do not agree for the scaling functions, surprisingly, the local pressure, temperature and density satisfy the virial equation of state for the hard sphere gas very well, except for a small deviation near the shock front  [see Fig.~\ref{fig:eqn_of_state}].  On the other hand, the radial velocity fluctuations are not gaussian, and is skewed towards positive fluctuations. This shows that local equilibrium is not attained. The hydrodynamic equations correspond to the collision less limit of the Boltzmann equation, and does not ensure equilibration. One way  to understand the role of these skewed distributions would be to study a system where the local velocities are reassigned at a constant rate from a Maxwell-Boltzmann distribution with the local temperature, and ask whether any qualitative changes are observed. This is a promising area for future study.

The divergence of temperature for small $\xi$ in the TvNS solution may be regularized by introducing heat conduction~\cite{afghoniem1982jfm,amraouf1991fluiddynres,hsteiner1994physfluids}. When the flow is made non-adiabatic, 
the conservation laws for mass and momentum do not change [see Eqs.~(\ref{eq:conservation-density}) and (\ref{eq:conservation-momentum})]. However, the conservation of energy [see Eq.~(\ref{eq:conservation-entropy})] is now modified to~\cite{Whithambook,afghoniem1982jfm,amraouf1991fluiddynres,hsteiner1994physfluids},
\be
 \rho \left[ \partial_t e + v \partial_r e \right] - T \left[ \partial_t \rho + v \partial_r \rho \right] + \frac{1}{r^2} \partial_r(r^2 q_r)=0,
 \label{eq:40}
\ee
where $e$ is the internal energy and $q_r$ is the heat flux in the radial direction, given by
\be
 q_r=-\lambda \nabla T =-\lambda\partial_r T,
\ee
where $\lambda$ is the  heat conductivity. Within kinetic theory, $\lambda$ depends on temperature as
\be
 \lambda=\sqrt{\frac{2 k_B^3 T}{\pi ^3 m \sigma^4}},
\ee
where $\sigma$ is the diameter of the particles and $m$ is the mass of the particles. In the limit $r,t \to \infty$, keeping $\xi$ fixed, it is easy to see that the ratio of the heat conduction term [last term in Eq.~(\ref{eq:40})] to any of the other terms decreases to zero as $t^{-2/5}$. Thus, this term is irrelevant in the scaling limit. However, one may take the limiting case of switching on a heat conduction term with a small coefficient, and determine the limit of the scaling functions as this coefficient tends to zero. The boundary condition of no heat flux at the heat center~\cite{afghoniem1982jfm,amraouf1991fluiddynres,hsteiner1994physfluids} automatically ensures that the gradients in temperatures are set to zero, or equivalently the scaling function $E \sim \xi^{-2}$, as seen in the simulations. 
Thus, the results for the scaling functions obtained by first taking the scaling limit at the level of the hydrodynamics equations, or by taking the scaling limit after solving the hydrodynamic equations with heat conduction may not be the same. Whether the latter  limit reproduces quantitatively the results of the simulations, requires a detailed numerical solution of the hydrodynamic equations, which is beyond the scope of this paper.

Earlier molecular dynamics simulations in two dimensions~\cite{mbarbier2016physfluids}  found that the simulations reproduce well the TvNS solution for low to medium densities, except for a small difference in the discontinuities at the shock front. Also, a slight discrepancy was observed near the shock center. When the number density of the ambient gas is high, the TvNS solution did not describe well the data near the shock front~\cite{mbarbier2016physfluids}. This is contrary to our results in three dimensions where the TvNS solution does not match with simulation. The number densities considered in both the studies are similar, and thus we expect the results in this paper to hold in two dimensions also. Ongoing work is focusing on understanding the reason for the qualitatively different results that have been reported for two and three dimensions.

\begin{acknowledgments}
The simulations were carried out on the supercomputer Nandadevi at The Institute of Mathematical Sciences.
\end{acknowledgments}

%\bibliography{refn}

\end{document}